\documentclass{PoS}

\usepackage{amsmath,amsfonts,amssymb,bbm,epsfig}

\title{The Tensor Pomeron and Small-x Deep Inelastic Scattering}

\ShortTitle{Tensor Pomeron and Small-x DIS}

\author{Daniel Britzger\\
        Max-Planck-Institut f\"ur Physik, F\"ohringer Ring 6, D-80805 M\"unchen, Germany\\
        E-mail: \email{britzger@mpp.mpg.de}}

\author{\speaker{Carlo Ewerz}\\
  Institut f\"ur Theoretische Physik, Universit\"at Heidelberg,\\
  Philosophenweg 16, D-69120 Heidelberg, Germany\\
  and\\
  ExtreMe Matter Institute EMMI, GSI Helmholtzzentrum f\"ur Schwerionenforschung,\\
  Planckstra{\ss}e 1, D-64291 Darmstadt, Germany\\
        E-mail: \email{C.Ewerz@thphys.uni-heidelberg.de}}

\author{Sasha Glazov, Stefan Schmitt\\
        Deutsches Elektronen-Synchrotron DESY, Notkestra{\ss}e 85, D-22607 Hamburg, Germany\\
        E-mail: \email{Alexandre.Glazov@desy.de}, \email{Stefan.Schmitt@desy.de}}

\author{Otto Nachtmann\\
        Institut f\"ur Theoretische Physik, Universit\"at Heidelberg,\\
Philosophenweg 16, D-69120 Heidelberg, Germany\\
        E-mail: \email{O.Nachtmann@thphys.uni-heidelberg.de}}

\abstract{
We apply the tensor-pomeron model to small-$x$ deep-inelastic
lepton-proton scattering and photoproduction. Our model includes
a soft and a hard tensor pomeron as well as a reggeon contribution. 
Data with c.\,m.\ energies $6 < \sqrt{s} < 318$ GeV and virtualities
$Q^2 < 50 \,\mbox{GeV}^2$ are considered. Our fit gives a very good 
description of the available data in this kinematic region, including
the latest HERA data for $x < 0.01$. In particular, the transition
region from low to high $Q^2$ is well described. Within the errors, the
hard pomeron is absent in photoproduction. The intercepts
of the soft and hard pomeron in the two-tensor-pomeron model
are found to be $1.0935\, ({}^{+76}_{-64})$ and 
$1.3008 \,({}^{+73}_{-84})$, respectively. We argue that a
vector pomeron would not give any contribution to photoproduction. 
}

\FullConference{XXVII International Workshop on Deep-Inelastic Scattering and Related Subjects - DIS2019\\
		8-12 April, 2019\\
		Torino, Italy}

\begin{document}

\section{Tensor pomeron}

High energy hadronic reactions are dominated by the physics of the pomeron.
For example, total cross sections, related to the elastic forward scattering amplitude
by the optical theorem, are at high c.\,m.\ energy described by the exchange of a pomeron
in the $t$-channel. For soft reactions, calculations of the pomeron from first principle
are currently not possible, and one has to retreat to Regge models to describe
soft high-energy scattering. Here, we want to discuss deep inelastic scattering (DIS)
in the context of such a Regge model. This presentation summarizes some of
the key results of \cite{Britzger:2019lvc} to which we refer the reader for further details. 

Until recently, the spin structure of the pomeron has not received much attention. 
It is well known that the pomeron carries vacuum quantum numbers with regard to
charge, color, isospin and charge conjugation. But what about spin? It has been 
shown some time ago that the pomeron can be regarded as a coherent superposition of
exchanges with spins 2, 4, 6, etc \cite{Nachtmann:1991ua}.
The new aspect that we want to discuss here is the
structure of the couplings of the pomeron to external particles. Most models treat
these couplings like those of a photon \cite{Donnachie:1983hf}, i.\,e., as vector
couplings. In many fits to DIS data the coupling is not specified at all and only
the energy dependence of the pomeron enters the formulae used for fitting the data,
see for example \cite{Donnachie:1998gm}. We argue that the pomeron couplings
play an important role, and that they should be treated as tensor couplings. 
Two of us, in collaboration with M.\ Maniatis, have constructed an effective
theory with such a tensor pomeron and reggeon contributions (and a vector odderon)
\cite{Ewerz:2013kda}. 

Conceptually, vector-type couplings of the pomeron turn out to be rather questionable. 
For example, a vector pomeron implies that the total cross sections for $pp$ and $\bar{p}p$
scattering at high energy have opposite sign. But, of course, quantum field theory
forbids negative cross sections. A further argument against a vector pomeron is that it 
does not give any contribution to photoproduction data, as we show in \cite{Britzger:2019lvc}. 
One may also ask about the possibility of a scalar coupling of the pomeron to external
particles. While possible from the point of view of quantum field theory, such
a coupling is experimentally disfavored. In \cite{Ewerz:2016onn} it was shown
that STAR data on polarized elastic $pp$ scattering are compatible with the tensor pomeron
but clearly rule out scalar pomeron couplings. 

\section{DIS in the tensor pomeron model}

Good fits of some DIS data in the context of Regge theory have been obtained in the literature,
see for example \cite{Donnachie:1998gm}. However, these do not explicitly treat the
couplings of the pomeron (assumed to be a vector) to the photon and proton.
In \cite{Britzger:2019lvc} we have addressed the question whether a Regge model
based on the tensor pomeron, treating in full detail its tensor 
couplings to the photon and proton, can successfully describe DIS and
photoproduction data. For making such a comparison to DIS data, we have added
a hard pomeron to the model of soft high-energy reactions of \cite{Ewerz:2013kda}.
The reggeon contribution, denoted by $\mathbbm{R}_+$, is dominated by the
$f_2$ but can contain a small contribution from the $a_2$. 
The parameters of the model are summarized in table \ref{tab1}. The indices
$j = 0,1,2$ always refer to the hard pomeron, the soft pomeron, and the reggeon,
respectively. 
\begin{table}[ht]\centering
\renewcommand{\arraystretch}{1}
\begin{tabular}{c|c|c|c}
& hard pomeron $\mathbbm{P}_0$ & soft pomeron $\mathbbm{P}_1$ & reggeon $\mathbbm{R}_+$ \\
\hline
intercept & $\alpha_0(0)=1+\epsilon_0$ & $\alpha_1(0)=1+\epsilon_1$ & $\alpha_2(0)=1+\epsilon_2$ \\
\hline
slope parameter & $\alpha'_0$ & $\alpha'_1$ & $\alpha'_2$ \\
\hline
$W^2$ parameter & $\tilde{\alpha}'_0$& $\tilde{\alpha}'_1$& $\tilde{\alpha}'_2$ \\
\hline
$pp$ coupling parameter & $\beta_{0pp}$ & $\beta_{1pp}$ & $\beta_{2pp}$ \\
\hline
$\gamma^* \gamma^*$ coupling functions & $\hat{a}_0(Q^2)$, $\hat{b}_0(Q^2)$ 
& $\hat{a}_1(Q^2)$, $\hat{b}_1(Q^2)$ & $\hat{a}_2(Q^2)$, $\hat{b}_2(Q^2)$ 
\end{tabular}
\caption{Parameters of our two-tensor-pomeron model. 
\label{tab1}}
\end{table}
The couplings of the pomerons and the reggeon to the photon contain functions
$\hat{a}_j$ and $\hat{b}_j$ which we cannot derive from first principles. In fitting
the data we make polynomial ans\"atze for them. 

We use the standard variables for DIS, namely the lepton-proton c.\,m.\ energy $\sqrt{s}$,
the photon virtuality $Q^2$, the c.\,m.\ energy $W$ of the photon-proton system, 
Bjorken $x=Q^2/(W^2+Q^2-m_p^2)$, and $y=(W^2+Q^2-m_p^2)/(s-m_p^2)$. 
Then, the total cross sections $\sigma_T$ and $\sigma_L$ for transversely and longitudinally
polarized photons resulting from our two-tensor-pomeron model are 
\begin{align}
\label{2.11}
&\sigma_T(W^2,Q^2)=4\pi \alpha_{\rm em} \, \frac{W^2-m_p^2}{W^2} \sum_{j=0,1,2} 
3 \beta_{jpp} (W^2 \tilde{\alpha}'_j)^{\epsilon_j} \cos\left(\frac{\pi}{2} \, \epsilon_j\right)
\\
& \:\: \times \left\{ 
\hat{b}_j(Q^2) \left[1+ \frac{2Q^2}{W^2-m_p^2} + \frac{Q^2 (Q^2+2m_p^2)}{(W^2 - m_p^2)^2}\right]
               - 2 Q^2 \hat{a}_j(Q^2) \left[1+ \frac{2Q^2}{W^2-m_p^2} + \frac{Q^2 (Q^2+m_p^2)}{(W^2 - m_p^2)^2} \right]\right\} ,
\nonumber \\
\label{2.12}
&\sigma_L(W^2,Q^2)=\, 4\pi \alpha_{\rm em} \, \frac{W^2-m_p^2}{W^2}\,Q^2 \sum_{j=0,1,2} 
3 \beta_{jpp} (W^2 \tilde{\alpha}'_j)^{\epsilon_j} \cos\left(\frac{\pi}{2} \, \epsilon_j\right)
\\
& \:\:\times \left\{ 
2 \hat{a}_j(Q^2) \left[1+ \frac{2Q^2}{W^2-m_p^2} + \frac{Q^2 (Q^2+m_p^2)}{(W^2 - m_p^2)^2}\right]
+ \hat{b}_j(Q^2) \frac{2m_p^2}{(W^2 - m_p^2)^2}
\right\} .
\nonumber
\end{align}
They are related to the structure functions $F_2$ and $F_L$ in the standard way, 
\begin{align}
\label{2.13}
F_2(W^2,Q^2) 
=& \, \frac{Q^2}{4\pi^2 \alpha_{\rm em}} (1-x) \left[1+2 \delta(W^2,Q^2)\right]^{-1} 
\left[\sigma_T (W^2,Q^2) + \sigma_L(W^2,Q^2) \right]
\\
\label{2.14}
F_L(W^2,Q^2) =& \, \frac{Q^2}{4\pi^2 \alpha_{\rm em}} (1-x) \sigma_L(W^2,Q^2) 
\end{align}
with $\delta(W^2,Q^2) = 2m_p^2Q^2/(W^2 + Q^2 -m_p^2)^2$. 
We use standard values for the electromagnetic coupling $\alpha_{\rm em}$ and the proton mass $m_p$. 

\section{Fit to DIS and photoproduction}

Using our model we perform a fit to the available DIS \cite{Abramowicz:2015mha} and
photoproduction data \cite{Aid:1995bz,Chekanov:2001gw,Vereshkov:2003cp,Caldwell:1978yb}
in the kinematic range given by $6 < \sqrt{s} < 318$ GeV, $Q^2 < 50 \,\mbox{GeV}^2$,
and $x<0.01$. We fit the reduced cross section 
\begin{equation}
\label{3.5}
\begin{split}
\sigma_{\rm red} (W^2,Q^2,y)=& \,\frac{1+(1-y)^2 + y^2 \delta(W^2,Q^2)}{1+(1-y)^2} 
\left[ F_2(W^2,Q^2) - \frac{\tilde{f}(W^2,Q^2,y) F_L(W^2,Q^2)}{1+2\delta(W^2,Q^2)} \right] \,,
\end{split}
\end{equation}
where 
\begin{equation}
\label{3.4}
\tilde{f}(W^2,Q^2,y) = \frac{y^2[1+2\delta(W^2,Q^2)]}{1+(1-y)^2 + y^2 \delta(W^2,Q^2)} \,.
\end{equation}
It contains the full experimental information and is the quantity that is actually measured. 
Some of the couplings and slope parameters in our model are well constrained from other
experiments so that we use their known values as default parameters.

Formally, our fit has 25 parameters. However, not all of them are equally important.
The most important parameters are the intercepts of the pomerons and of the reggeon. 
Further parameters of significance are the values of the coupling functions $\hat{a}_j(Q^2)$
and $\hat{b}_j(Q^2)$ at $Q^2=0$. The remaining parameters are used to describe the fall-off of 
the coupling functions $\hat{a}_j$ and $\hat{b}_j$ at large $Q^2$, and these 17 parameters
are much less significant for our result. The fit is of very satisfactory quality, for a detailed
quantification of the fit quality see \cite{Britzger:2019lvc}.
For the intercepts of the hard pomeron, the soft pomeron, and the reggeon we find, respectively, 
\begin{equation}
\alpha_0(0) = 1.3008 \,({}^{+73}_{-84})\,, \qquad
\alpha_1(0) = 1.0935 \,({}^{+76}_{-64})\,, \qquad
\alpha_2 (0) = 0.485 \, ({}^{+88}_{-90})\,.
\end{equation}

\begin{figure}
\begin{center}
  \includegraphics[width=0.5\textwidth]{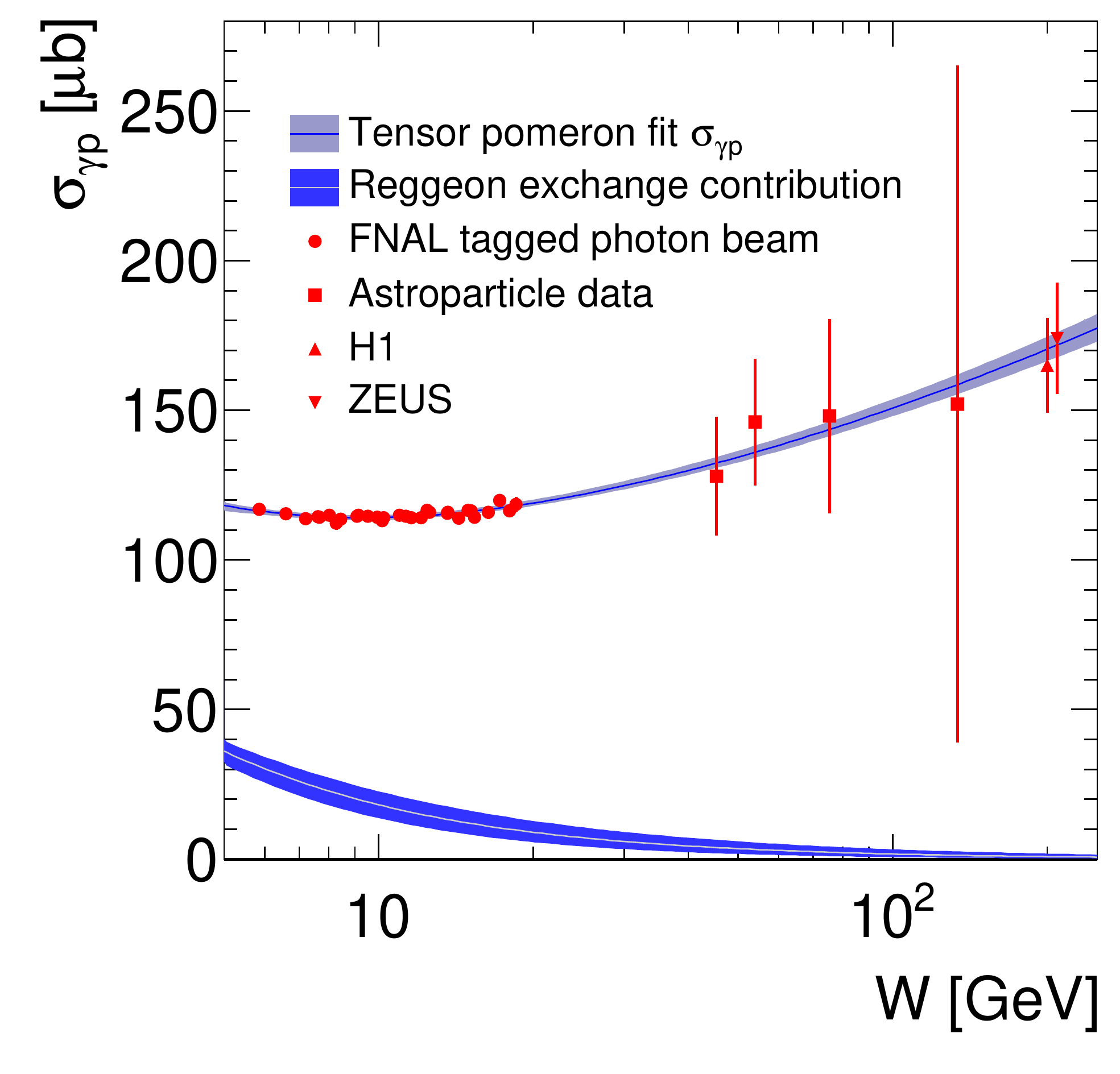}
  \caption{
    Comparison of our fit to the photoproduction cross sections 
    \cite{Aid:1995bz,Chekanov:2001gw,Vereshkov:2003cp,Caldwell:1978yb},
    with the reggeon contribution also shown separately. 
    The experimental uncertainties of the fit are indicated as shaded bands.
   \label{fig4}
   }
\label{fig:comp0}
\end{center}
\end{figure}

Figure \ref{fig:comp0} shows a comparison of our tensor-pomeron fit to the photoproduction
cross sections of \cite{Aid:1995bz,Chekanov:2001gw,Vereshkov:2003cp,Caldwell:1978yb}. 
We find that photoproduction is dominated by the soft pomeron, while the
hard pomeron contribution is compatible with zero here. As an example, we quote the
three different contributions for  $W=200$ GeV: 
\begin{equation}
\label{4.12c}
\begin{split}
170.4\, {}^{+4.2}_{-4.0} \: \mu{\rm b}& \qquad \mbox{for the soft pomeron } \mathbbm{P}_1\,,
\nonumber\\
0.002\, {}^{+0.086}_{-0.002} \: \mu{\rm b}& \qquad \mbox{for the hard pomeron } \mathbbm{P}_0\,,
\nonumber\\
0.84\, {}^{+0.99}_{-0.58} \: \mu{\rm b}& \qquad \mbox{for the $\mathbbm{R}_+$ reggeon}. 
\nonumber
\end{split}
\end{equation}
The reggeon contribution, also shown in the figure, is sizable for low $W$. We point out 
again that a vector pomeron would give zero contribution to the photoproduction cross section. 

\begin{figure}
\begin{center}
  \includegraphics[width=0.9\textwidth]{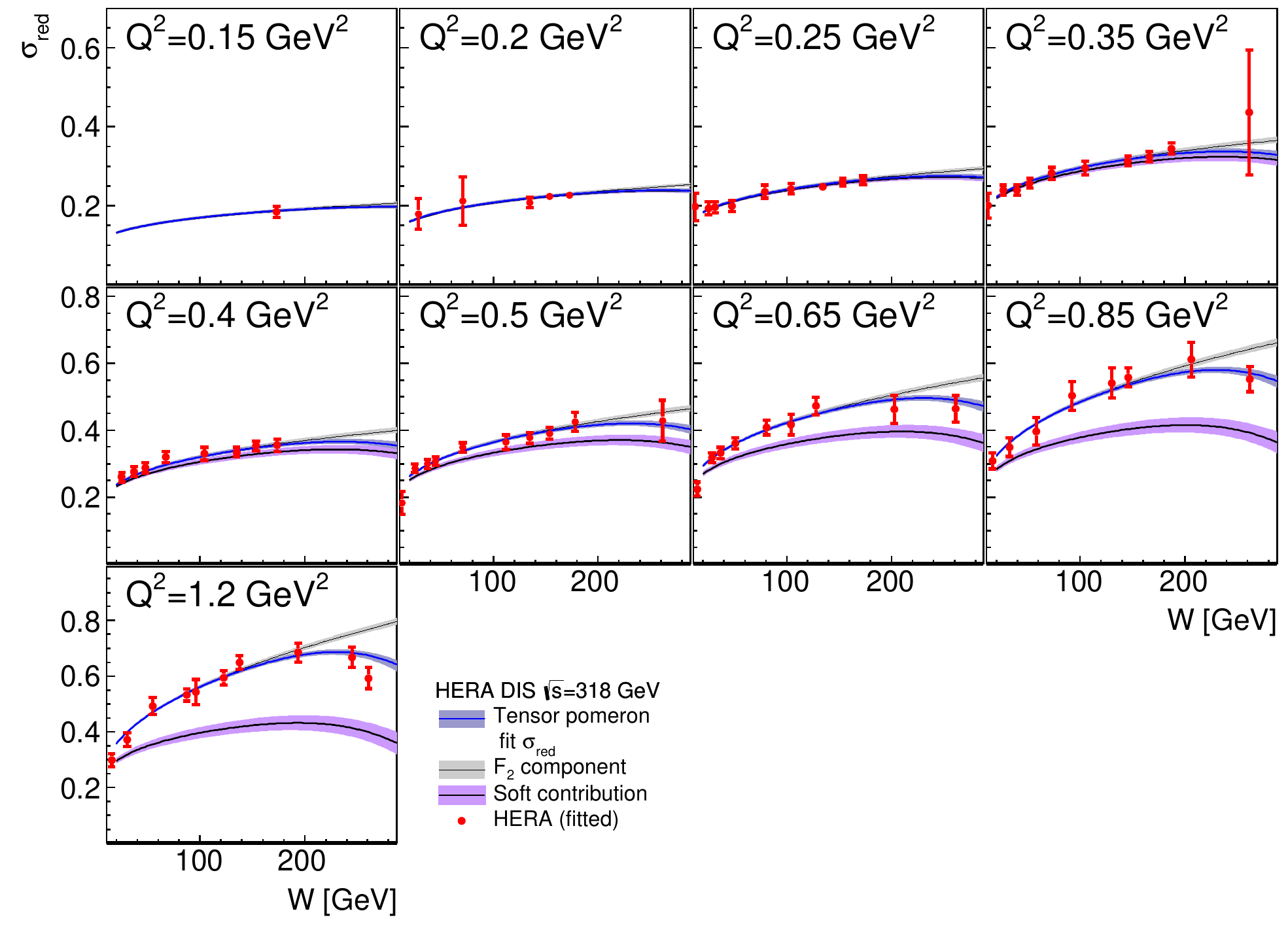}
  \caption{
    Comparison of our fit to DIS cross sections \cite{Abramowicz:2015mha} at $\sqrt{s}=318\,\text{GeV}$,
    at low $Q^2<1.5\,\text{GeV}^2$.
    We also show the soft contribution (soft pomeron plus $\mathbbm{R}_+$ reggeon) 
    and the contribution of the 
    structure function $F_2$ in the reduced cross section. 
    The experimental uncertainties of the fit are indicated as shaded bands.
\label{fig9} }
 \label{fig:comp4l}
\end{center}
\end{figure}

Next, we turn to DIS. Figure \ref{fig9} compares of our fit results, shown as a blue band, to the DIS
cross sections of \cite{Abramowicz:2015mha}. 
Here we have chosen as an example the data for $\sqrt{s}= 318$ GeV and low $Q^2<1.5\,\text{GeV}^2$.
We indicate the soft contribution (soft pomeron plus reggeon) as a purple band in figure \ref{fig9}.
The thin grey band represents the contribution of the structure function $F_2$ to
the reduced cross section, cf.\ equation \eqref{3.5}. 

The fit is of similarly good quality as in figure \ref{fig9} for the whole kinematic range that we consider. 
In the full kinematic range we find the following properties. 
The hard contribution increases with increasing $Q^2$, and the hard and soft contributions
are of approximately equal size at $Q^2 = 5\,\mbox{GeV}^2$.
But the soft contribution is still visible at $Q^2 = 20\,\mbox{GeV}^2$. 
The difference between $\sigma_{\rm red}$ and $F_2$, due to the longitudinal cross
section $\sigma_L$, is clearly visible at large $W$. As expected, the reggeon contribution becomes
very small at large $W$. 

We have also computed the ratio $R=\sigma_L/\sigma_T$ from our fit results for
the longitudinal and transverse cross sections. We compare this to the
$R$ data from H1 \cite{Andreev:2013vha} which are extracted directly from cross section
measurements at fixed $Q^2$ but different c.\,m.\ energies. 
We observe that our fit prefers higher values for $R$. The origin of this discrepancy
and its significance in view of the sizable error of the $R$ data from H1 are
difficult to assess. Measurements of $\sigma_L$ and of $R$ at a future electron-ion collider
would be very helpful to improve our understanding of the dynamics of the
longitudinal cross section. 

\section{Summary}

We have developed a two-tensor-pomeron model and have 
made a fit to photoproduction data and to small-$x$ DIS data from HERA. 
We obtain a very satisfactory fit and determine in particular the intercepts 
of the two pomerons. For DIS, the soft contribution 
is still clearly visible up to about $Q^2=20 \,\mbox{GeV}^2$. The transition 
from low to high $Q^2$ is nicely described. 

We have argued that a vector pomeron, that is a pomeron with vector type
couplings, is excluded by general arguments based on quantum field theory. 
For instance, it would not give any contribution to photoproduction
in clear contradiction with the data. 
A tensor pomeron, on the other hand, is compatible with quantum field
theory and permits an excellent description of photoproduction and DIS data,
as we have shown.

\end{document}